\documentclass[conference]{IEEEtran}
\IEEEoverridecommandlockouts
\usepackage{cite}

\usepackage[linesnumbered,vlined, ruled]{algorithm2e}
\usepackage{balance}
\usepackage{amsmath,amssymb,amsfonts}
\usepackage{mathtools,breqn}
\usepackage{graphicx}
\usepackage{textcomp}
\usepackage{xcolor}
\usepackage{hyperref}
\hypersetup{colorlinks,citecolor=black,urlcolor=black,linkcolor=black,unicode=true, breaklinks=false,pdfborder={0 0 1}}
\usepackage{enumerate}
\usepackage{subfig}
\usepackage{soul}

\SetKwRepeat{Do}{do}{while}
\usepackage[normalem]{ulem}

\newcommand{\Plus}{\raisebox{.1\height}{\scalebox{.6}{+}}}

\newcommand{\smallquotes}[1]

\definecolor{mygreen}{RGB}{0,150,0}

\SetCommentSty{mycommfont}

\SetKwProg{Pn}{Function}{:}{\KwRet}
\SetKwInOut{Input}{Input}
\SetKwInOut{Output}{Output}
\SetKwInOut{Procedure}{Procedure}
\SetKwFunction{TEST}{TEST}

\def\BibTeX{{\rm B\kern-.05em{\sc i\kern-.025em b}\kern-.08em
    T\kern-.1667em\lower.7ex\hbox{E}\kern-.125emX}}
\begin{document}

\title{Label Space Partition Selection for Multi-Object Tracking Using Two-Layer Partitioning
\thanks{$\ast$ Corresponding author}
}

\author{
    \IEEEauthorblockN{Ji Youn Lee$^{1}$, Changbeom Shim$^{1\ast}$, Hoa Van Nguyen$^{1}$,\\Tran Thien Dat Nguyen$^{1}$, Hyunjin Choi$^{2}$, and Youngho Kim$^{3}$}
    \IEEEauthorblockA{
        $^{1}$\textit{School of Electrical Engineering, Computing and Mathematical Sciences}, Curtin University, Perth, Australia\\
        $^{2}$\textit{AI R\&D Center}, AIBIZ Co. Ltd., Seoul, Republic of Korea\\
        $^{3}$CMW Geosciences, Perth, Australia\\
        jiyoun.lee2@postgrad.curtin.edu.au \; changbeom.shim@curtin.edu.au\; hoa.v.nguyen@curtin.edu.au\\t.nguyen1@curtin.edu.au \; hjchoi@ai-biz.net \; younghok@cmwgeo.com
    }
}

\maketitle

\begin{abstract}

    Estimating the trajectories of multi-objects poses a significant challenge due to data association ambiguity, which leads to a substantial increase in computational requirements. To address such problems, a divide-and-conquer manner has been employed with parallel computation. In this strategy, distinguished objects that have unique labels are grouped based on their statistical dependencies, the intersection of predicted measurements. Several geometry approaches have been used for label grouping since finding all intersected label pairs is clearly infeasible for large-scale tracking problems. This paper proposes an efficient implementation of label grouping for label-partitioned generalized labeled multi-Bernoulli filter framework using a secondary partitioning technique. This allows for parallel computation in the label graph indexing step, avoiding generating and eliminating duplicate comparisons. Additionally, we compare the performance of the proposed technique with several efficient spatial searching algorithms. The results demonstrate the superior performance of the proposed approach on large-scale data sets, enabling scalable trajectory estimation. 
\end{abstract}


\section{Introduction}
    The objective of Multi-Object Tracking (MOT) is to estimate the trajectories of an unknown number of objects varying over time from a series of sensor measurements. It is challenging due to the imperfections of the sensor generate uncertainties such as noisy measurements, false alarms, missed detections, and unknown data associations \cite{blackman1999design, bar2011tracking, mahler2007statistical, vo2014labeled}. In spite of these formidable challenges, MOT plays a pivotal role in diverse applications, including but not limited to surveillance~\cite{blackman1999design,bar2011tracking}, computer vision~\cite{cox1996an}, aerospace~\cite{reid1979algorithm}, cell biology~\cite{dat2021celltracking}, robotics~\cite{hoa2019jofr}. The majority of MOT algorithms can be categorized into three main frameworks: Multiple Hypothesis Tracking (MHT)~\cite{blackman1999design,bar2011tracking,reid1979algorithm}, Joint Probabilistic Data Association (JPDA)~\cite{bar2011tracking}, and Random Finite Sets (RFS)~\cite{mahler2007statistical}.
    
    The RFS-based framework, in particular, has gained popularity over the last two decades due to its rigorous mathematical foundations. It utilizes finite set-valued random variables to represent multi-object states and measurements and recursively propagates the multi-object density using the Bayes multi-object filter \cite{mahler2007statistical}, \cite{mahler2003multitarget}. Various RFS multi-object filters have been devised such as Probability Hypothesis Density (PHD) \cite{mahler2003multitarget}, Cardinalized Probability Hypothesis Density (CPHD) \cite{mahler2007phd}, Multi-Bernoulli (MB) \cite{mahler2007statistical, vo2008cardinality}, and Poisson Multi-Bernoulli Mixture (PMBM)~\cite{williams2015marginal} filters. However, the main drawback of these filters is their inability to estimate identities or trajectories, necessitating an additional post-processing procedure to reconstruct object trajectories. Notably, trajectories serve as a crucial tool for capturing object behavior, while labels facilitate the differentiation of individual trajectories and enable communication of relevant object information to both human and machine users. The first mathematically principled and tractable RFS-based multi-object tracker is the Generalized Labeled Multi-Bernoulli (GLMB) filter \cite{vo2013labeled,vo2014labeled}.  
    Its premise is the idea of labeled RFS which assigns a unique label to each object, thereby enabling the joint estimation of the object's states and trajectories.

    The GLMB filter has demonstrated its performance and availability \cite{vo2016efficient,vo2019multisensor,vo2019multiscan,nguyen2021tracking,van2021distributed, beard2020solution,trezza2022multi,moratuwage2022multi,shim2023linear}. In particular, remarkable scalability was shown by tracking over one million concurrent objects \cite{beard2020solution}. This achievement was made through the use of functional approximation and efficient parallel computation, providing evidence of the robustness and efficiency of the GLMB filter. A crucial aspect of the functional approximation involves computing a product of manageable and nearly independent smaller GLMB densities rather than relying on the large GLMB density itself. Specifically, smaller GLMB densities comprise label groups that have statistical dependence on each other and each label group is approximately independent.

    Although selecting a partition of the label space is feasible via \textit{label partitioning} \cite{beard2020solution}, further practical implementation of scalable MOT is still hindered by computationally inefficient procedures. Object groups well-separated in the measurement space exhibit low statistical dependence, as they seldom share the same measurements. Especially in the measurement update stage, it is notably time-consuming within the scalable GLMB filter, as it necessitates the examination of all potential label pairs to ascertain whether their gating regions intersect. This bottleneck becomes particularly pronounced in large-scale and/or high-density multi-object scenarios. Therefore, the need for an efficient label partitioning implementation is imperative to enable the practical use of scalable GLMB filters in real-world scenarios.

    To improve computational efficiency in label partitioning, spatial indexing techniques can be leveraged. One frequently employed method is data-driven partitioning, where objects are approximated in each different level of segment-tree like R-tree\cite{guttman1984r} and its variants\cite{beckmann1990r, sellis1987r+, kim2001optimizing}. In \cite{beard2020solution}, R-tree was utilized for indexing labels and examining whether their gating regions intersect spatially. On the other hand, space-driven partitioning involves dividing the space into spatially disjoint partitions such as kd-tree \cite{bentley1975multidimensional}, quad-tree \cite{finkel1974quad}, and grid \cite{bentley1979data}. Among these, exploiting the grid structure supports straightforward implementation and parallelization, making it suitable for in-memory processing \cite{mouratidis2005conceptual,demiryurek2009efficient,kim2019moving,ijcai2023p0618}, which was also studied for label partitioning \cite{shim2021space, shim2022generalized}. However, a potential drawback is to get rid of excessive duplicates when retrieving intersecting objects in a dense area due to objects being present in multiple grid cells \cite{jacox2007spatial}. Numerous approaches have been introduced to address the issue of duplicate results in space-driven partitioning \cite{aref1994hashing, patel1996partition, zhou1998data, dittrich2000data, yu2019spatial}.

    In this paper, we propose an efficient implementation for label space partition selection in label-partitioned GLMB filtering. The computational bottleneck for parallel MOT is mitigated by using the latest space-oriented technique. Specifically, we employ a two-layer partitioning technique \cite{tsitsigkos2021two} when searching statistically dependent objects, which is important for grouping multiple objects and processing parallel GLMB filtering. To reduce the number of intersection tests, every object within each spatial partition is further divided into secondary classifying. In our implementation, redundant steps in checking overlapping labels are pruned so that a feasible label space partition can be found efficiently. A numerical study is conducted to evaluate the performance of the proposed method.

    The remainder of this paper is organized as follows. Section \ref{sec:preliminaries} gives essential background on the GLMB filters, and Section \ref{sec:propose} presents our proposed approach. The performance of our method is evaluated in section \ref{sec:experiments} through various experiments. Finally, Section \ref{sec:conclusion} provides concluding remarks.

\section{Preliminaries}\label{sec:preliminaries}
    This section provides a brief background for this paper. Initially, in Section \ref{subsec:GLMB}, we introduce the standard GLMB filter. Section \ref{subsec:LGLMB} delves into the scalable GLMB filter and label partitioning. Additionally, to provide insight from a spatial join perspective, we introduce prominent spatial join techniques in Section \ref{subsec:spatial-join}. For a more comprehensive understanding, further details can be found in references \cite{vo2013labeled, vo2014labeled, beard2020solution, jacox2007spatial}.
    \subsection{Generalized Labeled Multi-Bernoulli Tracker}\label{subsec:GLMB}
        The GLMB filter is designed to estimate object trajectories by modeling the multi-object state with labeled RFS. In particular, a multi-object state $\textbf{X}$ is a collection of the single-object state $\textbf{x} = (x,\ell)$  where $x\in\mathbb{X}$ is its kinematic state, and $\ell$ is a distinct label in some discrete label space $\mathbb{L}$. We define a projection $\mathcal{L}: \mathbb{X} \times \mathbb{L} \rightarrow \mathbb{L}$ by $\mathcal{L}((x, \ell)) = \ell$. A set $\textbf{X}$ of labeled objects has no duplicate labels if $\textbf{X}$ and its labels $\mathcal{L}(\textbf{X}) = \{\mathcal{L}(\textbf{x}) : \textbf{x} \in \textbf{X}\}$ have the same cardinality, i.e., $|\mathcal{L}(\textbf{X})| = |\textbf{X}|.$          
        In other words, if the labels are distinct we have $\Delta(\textbf{X}) \triangleq \delta _{|\textbf{X}|}(|\mathcal{L}(\textbf{X})|) = 1,$
        where the generalized Kronecker delta function is defined as
            $$\delta _{Y}(X) \triangleq  \begin{cases}
             1, & \text{if} ~~ X = Y\\
             0, & \text{otherwise} 
            \end{cases}, $$ (with arbitrary argument, i.e., set, vector or scalar).
        Besides, the integral of a function $\textbf{f}:\mathcal{F}(\mathbb{X}\times \mathbb{L}) \to \mathbb{R}$ is given by $\int \textbf{f}(\textbf{x})d\textbf{x} = \sum_{\ell \in \mathbb{L}} \int_{\mathbb{X}} \textbf{f}((x,\ell))d x.$
    
        For compact expression, following \cite{vo2013labeled}, we define the multi-object exponential as
            $$\vartheta^X \triangleq \prod_{x\in X} \vartheta(x),$$
        and the set inclusion function follows as         
        $$ 1_{Y}(X) \triangleq \begin{cases}
              1, & \text{if} ~~ X \subseteq Y \\
              0, & \text{otherwise}
             \end{cases}. $$

        The GLMB filter propagates the GLMB (multi-object) density of the form
        \begin{equation}
            \label{eqn:GLMBdensity}
                \boldsymbol{\pi}(\textbf{X}) = \Delta(\textbf{X})\sum_{\xi \in \Xi}{w^{(\xi)}(\mathcal{L}(\textbf{X}))\Big[p^{(\xi)}\Big]^{\textbf{X}}},
            \end{equation}
        where: $\Xi$ is a discrete index set; $p^{(\xi)}(\cdot,\ell)$ is a probability density; and $w^{(\xi)}(I)$ is non-negative weight that satisfies $\sum_{\xi \in \Xi}\sum_{L \in \mathcal{F}(\mathbb{L})}w^{(\xi)}(L) = 1$ with $\mathcal{F}(\mathbb{L})$ is all finite subsets of $\mathbb{L}$ including the empty set. The GLMB filter is a closed-form multi-object tracking filter since the prior and the filtering densities are both GLMB families. At each time step, the number of terms in the filtering GLMB density grows exponentially, hence only significant components (ones with high weights) are retained to meet the computational constraint.
    
    \subsection{Label-partitioned GLMB Filtering}\label{subsec:LGLMB}
        In the standard GLMB filter implementation, the computational demand grows significantly when the number of objects increases since the filter complexity is at least in the square number of objects\cite{vo2017efficient}. The primary computational bottleneck occurs during the measurement update stage, with statistical dependence mainly arising from uncertainty in the associations between measurements and objects. This challenge is especially pronounced in dense scenarios where objects are close to each other.
        
        Nevertheless, in practice, objects often travel in groups (i.e., pedestrians, flock of birds), where each group can be assumed to be well-separated from the other. Exploiting this fact, a scalable implementation of the GLMB filter \cite{beard2020solution} approximates the GLMB density by a product of GLMBs for relatively independent groups of objects. It reduces the complexity to the logarithmic magnitude in the number of objects from the quadratic magnitude in the standard implementation.
        
        Let $\mathfrak{L}$ be some partition of the label space $\mathbb{L}$. A labeled RFS density on $\mathcal{F}(\mathbb{X}\times\mathbb{L})$ is considered as $\mathfrak{L}$ \textit{-partitioned GLMB density} $\boldsymbol{\pi}_{\mathfrak{L}}$ if it can be expressed as a product of GLMBs (each defined on $\mathcal{F}(\mathbb{X}\times L)$,  for $L\in\mathfrak{L}$), as follows
        \begin{align}
          \boldsymbol{\pi}_{\mathfrak{L}}(\boldsymbol{X}) = \prod_{L\in \mathfrak{L}} \boldsymbol{\pi}_{\mathfrak{L}}^{(L)} (\boldsymbol{X} \cap (\mathbb{X} \times L)).  
        \end{align}
        If the current filtering density is an $\mathfrak{L}$-partitioned GLMB, the new filtering density $\boldsymbol{\pi}_{\Plus}$ at the next time step is a standard GLMB \cite{beard2020solution}, not suitable for scalable implementation. To maintain the scalability, the new filtering density can be approximated by an $\mathfrak{L}$-partitioned GLMB. Nevertheless, since the current partition $\mathfrak{L}$ might not be a suitable partition for approximating the new filtering density,  the goal is to find the optimal partition $\mathfrak{L}_{\Plus}$ of the next time step label space $\mathbb{L}_{\Plus}$ (in some statistical sense) in the space of all possible partitions of $\mathbb{L}_{\Plus}$ (denoted as $\mathcal{P}(\mathbb{L}_{\Plus})$) for the new filtering density.
        
    \subsection{Label Partitioning}\label{subsec:spatial-join} 
        Label partitioning in the scalable GLMB filter \cite{beard2020solution} aims to find a partition of the labels to minimize the amount of potential measurement sharing between objects in different groups. Naturally, labels that are well-separated in the measurement space exhibit low statistical dependence because the likelihood of these labels producing closely spaced measurements is minimal. Here, ``well-separated" means the distance between trajectories of labels is significantly larger than measurement noise and predicted object position uncertainty.
        Assume each factor $\boldsymbol{\pi}^{(L)}_{\mathfrak{L}}$ has the following form:
        \begin{align}
            \boldsymbol{\pi}_{\mathfrak{L}}^{(L)} =  \left\{\begin{pmatrix}w_{\mathfrak{L},L}^{(I,\xi)}, p_{\mathfrak{L},L}^{(\xi)}\end{pmatrix}\right\}_{(I,\xi) \in \mathcal{F}(L)\times\Xi^{(L)}}.
        \end{align}
        Then at the next time step, the distribution of the predicted measurement of  each label $\ell \in \mathbb{L}_{\Plus}$ is given by
        \begin{align}
            \tilde{p}_{\Plus}(z,\ell) = \int g(z|x,\ell)p_{\Plus}(x,\ell) dx    
        \end{align}
        where $p_{\Plus}(x,\ell)$ is the predicted state and $g(z|x,\ell)$ is the single-object measurement likelihood function. 

        For each label $\ell \in \mathbb{L}_{\Plus}$, one can use the distribution $\tilde{p}_{\Plus}(\cdot,\ell)$ to establish a gating region, $B(\ell) \subseteq \mathbb{Z}$, which holds most of the probability mass for the predicted measurement.
        These gating regions form the foundation for dividing the new label space $\mathbb{L}_{\Plus}$ into a partition $\mathfrak{L}_{\Plus} = \{L_1,\dots,L_{|\mathfrak{L}_{\Plus}|} \}$. Furthermore, to find feasible label partitions, the maximum number of labels within a single partition is constrained by $L_{\text{max}}$, i.e., $|L|\leq L_{\text{max}}, \forall L \in \mathfrak{L}_{\Plus}$. In summary, $\mathfrak{L}_{\Plus}$ must meet the following conditions~\cite{beard2020solution}:  
        
        \begin{enumerate}
            \item For all $L \in \mathfrak{L}_{\Plus}$ and for any $\ell_i, \ell_j \in L$, either $B(\ell_i) \cap B(\ell_j) \neq \emptyset$ or $B(\ell_i) \cap B(\ell_1) \neq \emptyset$, $B(\ell_1) \cap B(\ell_2) \neq \emptyset$, ..., $B(\ell_n) \cap B(\ell_j) \neq \emptyset$ for $\{\ell_1, ..., \ell_n\} \subseteq L$;
    
            \item For all $i,j \in \{1,...,|\mathfrak{L}_{\Plus}|\}$ and $i \neq j$, $\big[\bigcup_{\ell \in L_i}B(\ell)\big] \cap \big[\bigcup_{\ell \in L_j}B(\ell)\big] = \emptyset$; and
            \item For all $L \in \mathfrak{L}_{\Plus}$, $|L| \leq L_{max}$.
        \end{enumerate}
        The process involves updating entire prediction gates to decrease the gating probability $P_G$ until the specified limit of $L_{\text{max}}$ is reached. Algorithm \ref{algo:label_partitioning} illustrates the proposed label partitioning process.

            \begin{algorithm}[t!]
                \caption{Label Space Partitioning \cite{beard2020solution}}
                \label{algo:label_partitioning}
                \SetKwData{Up}{up}
                \SetKwFunction{BuildGrid}{BuildGrid}
                \SetKwFunction{FindGMBR}{FindGMBR}
                \SetKwFunction{MakeLabelGraph}{MakeLabelGraph}
                \SetKwFunction{GeneralLabelGraph}{GeneralLabelGraph}
                \SetKwFunction{GroupLabel}{GroupLabel}
                \SetKwFunction{GetEncloseRect}{GetEncloseRect}
                \SetKwFunction{DFS}{DFS}
                \SetKwFunction{SelectLabelPartition}{SelectLabelPartition}
                
                \SetKwProg{Pn}{Function}{:}{\KwRet}
                
                \SetKwInOut{Input}{Input}
                \SetKwInOut{Output}{Output}
                \SetKwInOut{Procedure}{Procedure}
                \Input{$\mathcal{L}(\textbf{X}), P_G, L_{\text{max}}$}
                \Output{$\mathfrak{L_{\Plus}}$}
                \Do{$\max|L|$ $\leq$ $L_{\text{max}}$}{
                   $M \leftarrow \texttt{FindGMBR(} \mathcal{L}(\textbf{X}), P_G \texttt{)}$\;
                   $\mathfrak{P} \leftarrow \texttt{SelectLabelPartition(} M \texttt{)}$\;
                   $\mathfrak{L_{\Plus}} \leftarrow \texttt{GroupLabel(} \mathfrak{P} \texttt{)}$\;

                   Reduce $P_G$\;
                }
            \end{algorithm}
    
        Finding $\mathfrak{L}_{\Plus}$ poses challenges regarding time complexity due to the necessity of examining all possible label pairs to determine if their predicted gating regions $B(\ell)$, referred to as Gaussian-mixed Minimum Bounding Rectangles (GMBRs), intersect. Additionally, this step must be reiterated until each partition's maximum label count is no greater than $L_{\text{max}}$. In \cite{beard2020solution}, an R-tree is employed to alleviate the computational complexity of the partitioning, making it feasible for such problems. However, using an R-tree structure for indexing becomes increasingly burdensome as the number of procedure iterations grows~\cite{tsitsigkos2021two}. 
        
        To address these computational challenges, an alternative approach is to employ a grid structure for object indexing and retrieval. Indexing the object with its associated grid can be accomplished with linear time complexity, through algebraic operations \cite{vsidlauskas2009trees, ray2014supporting}. While indexing objects offers computational advantages, the issue of duplicate results arises when objects are assigned to multiple grid-based tiles, necessitating additional processing. 
        
        A straightforward approach to eliminate duplicates is by hashing the search results and checking for duplicates within each candidate set \cite{aref1994hashing}. While this method is simple, it can be costly, particularly when dealing with a large number of results. Another widely used method involves using a reference point within the intersection area between each result and the search range. If the reference point falls inside the partition, the result is reported; otherwise, it is discarded \cite{dittrich2000data}. While this method eliminates the need for hashing, we must account for the cost associated with identifying duplicate results and computing the reference point for each of them  \cite{tsitsigkos2021two}.

\section{Efficient Label-partitioning for GLMB Filters}\label{sec:propose}
    A secondary partitioning approach refers to a spatial indexing technique used in \textit{space-driven partitioning} spatial index, such as a grid. It involves categorizing the indexed objects as rectangles within each spatial partition into four secondary classes, as introduced in \cite{tsitsigkos2021two}. This approach reduces the number of comparisons needed during intersection assessments and eliminates the generation of duplicate results on the grid. 
    
    Section \ref{subsec:secandary} provides an overview of the secondary partitioning technique proposed in \cite{tsitsigkos2021two}. Section \ref{subsec:implementation} describes the approach for label partitioning using this secondary partitioning technique. 

    \subsection{Secondary Classifying for Label Partitioning}\label{subsec:secandary}
        \subsubsection{Secondary Classifying}\label{subsubsec:sec_classifying}
        Consider a regular grid that divides the space into $N \times M$ disjoint spatial partitions, referred to as tiles. Given a set of GMBRs, each GMBR $r$ is indexed into tiles that spatially intersect with $r$. Assume a 2D dimension, i.e., dimension $d \in \{x,y\}$, the coordinates of the GMBR are denoted as intervals; $r.x = \left[ r.x_l, r.x_u \right ]$  and $r.y = \left[ r.y_l, r.y_u \right ]$ on the respective $x$ and $y$ axes. Here, $r.d_l$ and $r.d_u$ denote the minimum and maximum coordinates of the GMBR $r$ on each dimension, respectively. Subsequently, each GMBR $r$ assigned to each tile $T$ is further classified into a secondary class, $\mathfrak{A}$, $\mathfrak{B}$, $\mathfrak{C}$ and $\mathfrak{D}$, as follows:
        
        \begin{itemize}
          \item $r$ is categorized into $\mathfrak{A}$, if $r$ originates within tile $T$ for both dimensions, i.e., 
                \begin{description}
                    \item if, $T.x_l \leq r.x_l$ and $T.y_l \leq r.y_l$;
                \end{description}
          \item $r$ is categorized into $\mathfrak{B}$, if $r.x$ starts within $T.x$ and $r.y$ starts before $T.y$, i.e., 
                \begin{description}
                    \item if, $T.x_l \leq r.x_u$ and $T.y_l > r.y_l$;
                \end{description}
          \item $r$ is categorized into $\mathfrak{C}$, if $r.x$ starts before $T.x$ and $r.y$ starts within $T.y$, i.e., 
                \begin{description}
                    \item if, $T.x_l > r.x_l$ and $T.y_l \leq r.y_l$; and
                \end{description}
          \item $r$ is categorized into $\mathfrak{D}$, if $r$ originates before tile $T$ for both dimensions, i.e., 
                \begin{description}
                    \item if, $T.x_l > r.x_l$ and $T.y_l > r.y_l$.
                \end{description}
        \end{itemize}
            \begin{figure}[t!]
                \centering
                    \subfloat[Secondary Classes]{{\includegraphics[width=7.50cm]{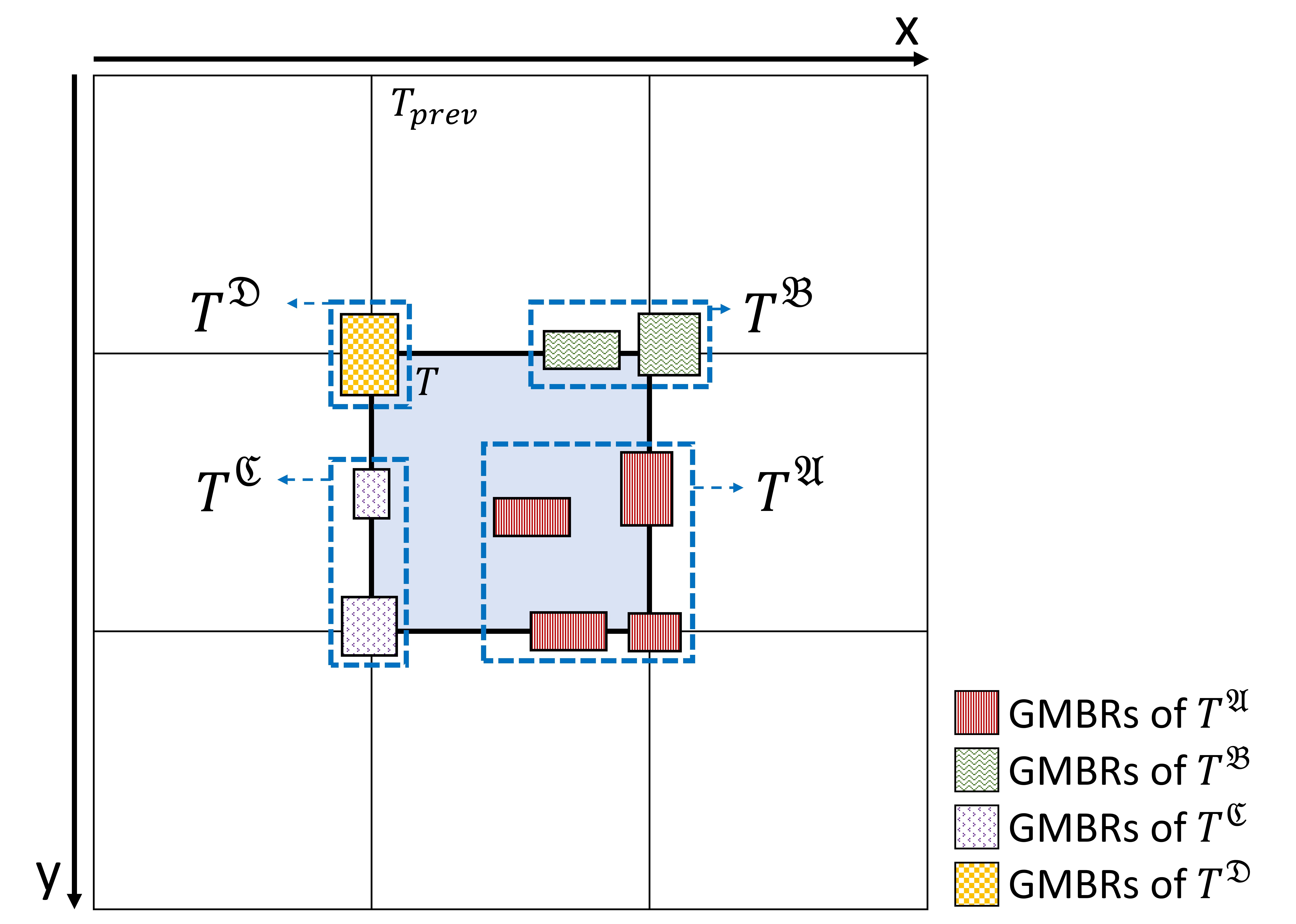} }}\hspace{-0cm}\vspace{-0.3cm}
                    \subfloat[Selecting Relevant Classes]{{\includegraphics[width=7.70cm]{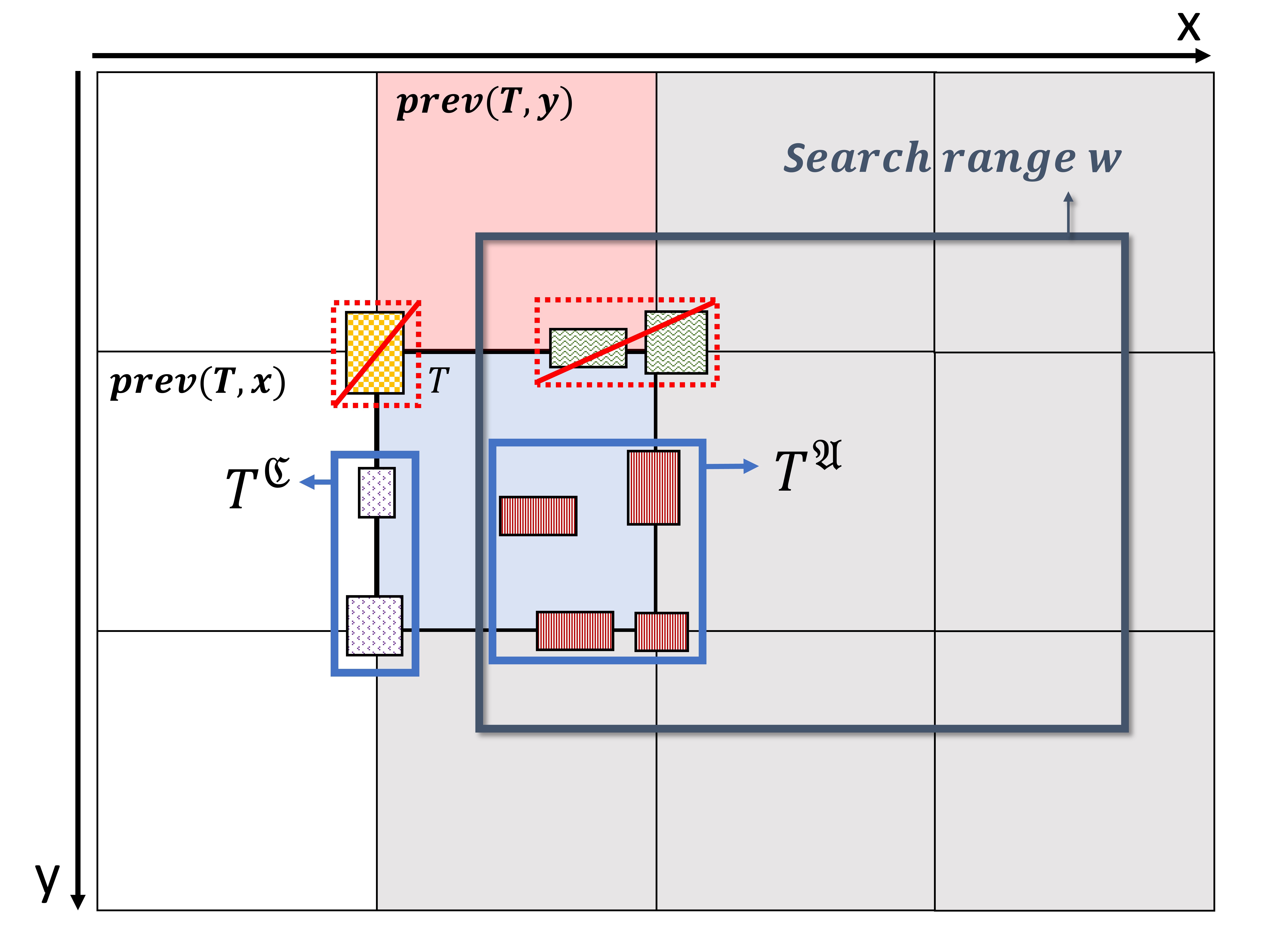} }}
                \caption{Examples of Secondary Classifying}
                \label{fig:sec_class}
                \vspace{-0.2cm}
            \end{figure}
            
        Fig.~\ref{fig:sec_class}~(a) provides visual examples of GMBRs classifying into the four secondary classes within a tile. For convenience, the secondary classes of tile $T$ are denoted as $T^{\mathfrak{A}}$, $T^{\mathfrak{B}}$, $T^{\mathfrak{C}}$ and $T^{\mathfrak{D}}$ respectively.
        
        \subsubsection{Selecting Valid Classes}\label{subsubsec:selecting_classes}
            Label partitioning utilizing a grid structure can yield duplicated results when attempting to find groups of intersecting GMBRs. For instance, in Fig.~\ref{fig:sec_class}~(a), rectangles within both $T^{\mathfrak{B}}$ and $T^{\mathfrak{D}}$ are compared twice: once when inspecting the tile $T_{prev}$ and then again when inspecting tile $T$. Consequently, results may be duplicated if intersecting GMBRs are present. Given a search range $w$, secondary classes prevent the generation of duplicate results based on the following criteria: 
              \begin{itemize}
                \item If $w$ intersects with tiles $T$ and starts before tile $T$ in the $x$ dimension, then it is necessary to exclude $T^{\mathfrak{C}}$ and $T^{\mathfrak{D}}$;
                \item If $w$ intersects tile $T$ and $w$ starts before $T$ in dimension $y$, then $T^{\mathfrak{B}}$ and $T^{\mathfrak{D}}$ should be disregarded; and
                \item If $w$ intersects tile $T$ and $w$ starts before $T$ in both dimensions, then only $T^{\mathfrak{A}}$ should be considered.
              \end{itemize}
          Note that the proof of these criteria can be found in \cite{tsitsigkos2021two}. 

           Fig.~\ref{fig:sec_class}~(b) provides an illustrative example of selecting relevant classes. Let $prev(T, d)$ denote the tile that precedes $T$ in the specified dimension $d$. Given a search range $w$, only the tiles that intersect with $w$ are sequentially inspected. In this scenario, $w$ starts before tile $T$ only in the $y$ dimension. Therefore, we can selectively check the rectangles in the $T^{\mathfrak{A}}$ and $T^{\mathfrak{C}}$ because rectangles in the $T^{\mathfrak{B}}$ and $T^{\mathfrak{D}}$ have been examined when $prev(T,y)$ was inspected. Note that more stages for reducing the number of comparisons have been proposed in \cite{tsitsigkos2021two}. However, this work does not consider these additional criteria because they may not be suitable here.

        \subsection{Implementation of Label Partitioning}\label{subsec:implementation}
        We proposed a computationally efficient method for label space partitioning combined with a secondary classifying approach. It possesses a notable advantage in terms of ease of implementation and seamless integration into diverse, complex research projects and large-scale spatial data management systems, facilitating parallel computation. Moreover, it effectively lowers time while consistently delivering robust performance. While this method is applicable to any \textit{space-driven partitioning} spatial index, we have specifically tailored it for use within a two-dimensional grid structure to optimize label partitioning efficiency. 

        \begin{algorithm}[t!]
                \caption{Label Space Partition Selection}\label{alg:range_query}
                    \SetKwProg{Pn}{Function}{:}{\KwRet}
                    \SetKwFunction{TwoLayerLabelPartition}{TwoLayerLabelPartition}
                    \SetKwFunction{SecondaryClassifying}{SecondaryClassifying}
                    \SetKwFunction{GetRelevantTiles}{GetRelevantTiles}
                    \SetKwFunction{GetRelevantClasses}{GetRelevantClasses}
                    \SetKwFunction{IntersectingTest}{IntersectingTest}
                    \SetKwFunction{MakeLabelGraph}{MakeLabelGraph}
                    \KwIn{Grid $\mathcal{G}$, GMBRs $M$}
                    \KwOut{Label partitions $\mathfrak{P}$}
        
                    \Pn{\TwoLayerLabelPartition{$\mathcal{G}, M$}}{
                        $\mathfrak{P} \leftarrow \emptyset$\;
                        $\mathcal{S}_{\mathcal{G}} \leftarrow$ \SecondaryClassifying{$\mathcal{G}, M$}\;
                        \ForEach{GMBR $w \in M$}{
                            $T_m \leftarrow$ \GetRelevantTiles{$\mathcal{G}, w$}\;
                            $\mathcal{S}_{t} \leftarrow$    \GetRelevantClasses{$w, T_m, \mathcal{S}_{\mathcal{G}}$}\;
                            $\mathcal{I}_w \leftarrow$ \IntersectingTest{$w$, $T_w$, $\mathcal{S}^{X}_t$}\;
                        $\mathfrak{P} \leftarrow \mathfrak{P} \cup (w, \mathcal{I}_w)$\;
                        }

                    \Return $\mathfrak{P}$;
                    }    
            \end{algorithm}

            Algorithm~\ref{alg:range_query} describes the label space partition selection of label partitioning for scalable GLMB filters. Given a set of GMBRs $M$ and grid configuration $\mathcal{G}$, all GMBRs are classified into secondary classes for each tile $t$ in the \texttt{SecondaryClassifying} stage based on \ref{subsubsec:sec_classifying}. Following this, the intersection search process begins to find spatially intersecting GMBRs within a specified search range. Here, each GMBR from the set $M$ is used as a search range $w$. During each iteration, relevant tiles intersecting $w$ are selected using algebraic operations in the \texttt{GetRelevantTiles} stage. Within the relevant tiles containing information about the secondary classes, the \texttt{GetSecondaryClass} stage is executed to select the most relevant classes for each tile based on \ref{subsubsec:selecting_classes}. By selecting the relevant secondary classes, \texttt{IntersectingTest} stage finds intersected GMBRs with $w$ among a set of GMBRs from these classes. As a result, the \texttt{TwoLayerLabelPartition} function returns label partitions that group GMBRs intersecting with each other. In summary, the secondary classifying technique reduces redundant comparisons, leading to computational efficiency in the label partitioning function.
        
\section{Numerical Study}\label{sec:experiments}
    In this section, we discuss the experimental performance of our method. The effectiveness and computational efficiency are evaluated by changing the number of labels. All experiments were conducted on a machine with 64GB of RAM and an Intel(R) Xeon(R) CPU E5-2696 v3 @ 2.30GHz running Ubuntu 22.04 
    using Python 3.10.12.

    \vspace{0.15cm}\noindent\textbf{Baseline.} 
    We evaluate the simulation results by comparing the proposed method with secondary partitioning (so-called Two-layer) to two existing approaches: R-tree \cite{beard2020solution} and \textit{inclusion-checking grid} (IG) \cite{shim2021space}. IG utilizes grid-based label partitioning along with a pruning mechanism to decrease the number of intersection checks and time complexity. Our approach adopts the principles of IG but goes a step further by combining secondary classification to eliminate duplicated results. Note that all filtering processes are the same across the three methods, with only the label partitioning stage being adjusted. The results of label partitioning in all three methods are identical, meaning they correctly determine intersected GMBRs for each GMBR, although with variations in processing time.

    \vspace{0.15cm}\noindent\textbf{Data sets.}
    We employed diverse datasets encompassing varying object cardinalities, ranging from 10K to 100K. These objects were randomly generated with a uniform distribution and represented by GMBRs with a maximum width or height of 20m. Furthermore, our analysis focused on a 2 km $\times$ 2 km surveillance area and the increase of cardinalities indicates that the test environment is going to be denser. By default, we configured the tests with 30K objects and 100 tiles on each axis of the grid for other experiments.\

    \begin{figure}%
    \centering
        \includegraphics[width=9.0cm]{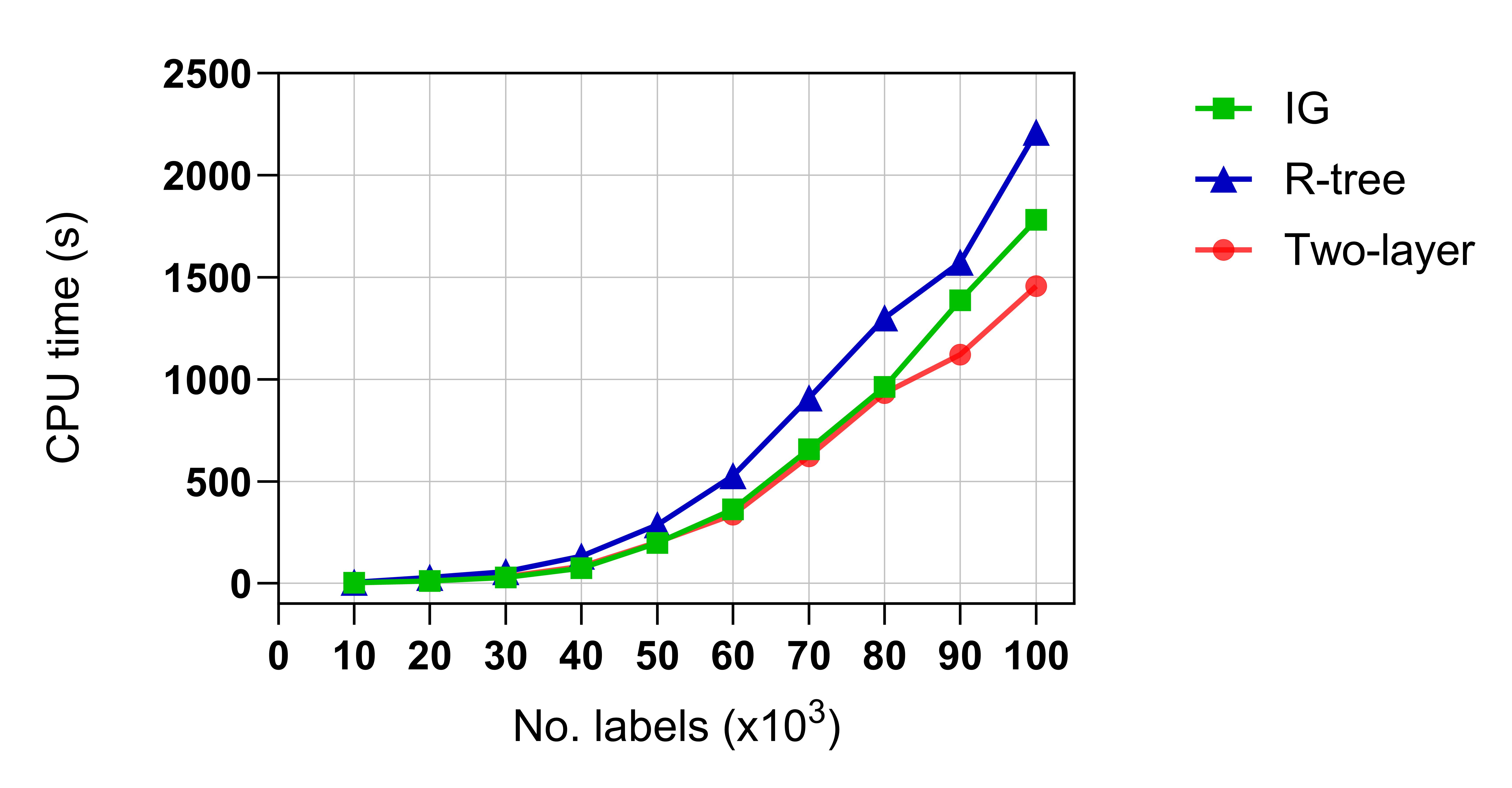}
        \caption{Comparison of the total processing time}
        \vspace{-0.4cm}
        \label{fig:total_cpu_time}%
    \end{figure}

    The initial experiments involve comparing the performance of the proposed approach with that of R-tree and IG across datasets of varying sizes. In Fig.~\ref{fig:total_cpu_time}, we present the total CPU time results for different numbers of labels. It is evident that the two grid-based structures consistently outperform the R-tree in terms of CPU times. Furthermore, the proposed algorithm performs similarly to IG until the dataset size reaches 50K objects. However, it surpasses IG thereafter, and the performance gap widens as the dataset size increases. The experimental results emphasize the effective reduction in time complexity achieved by the proposed method, particularly in dense environments. 
    
\begin{figure}%
        \centering
        \includegraphics[width=9.0cm]{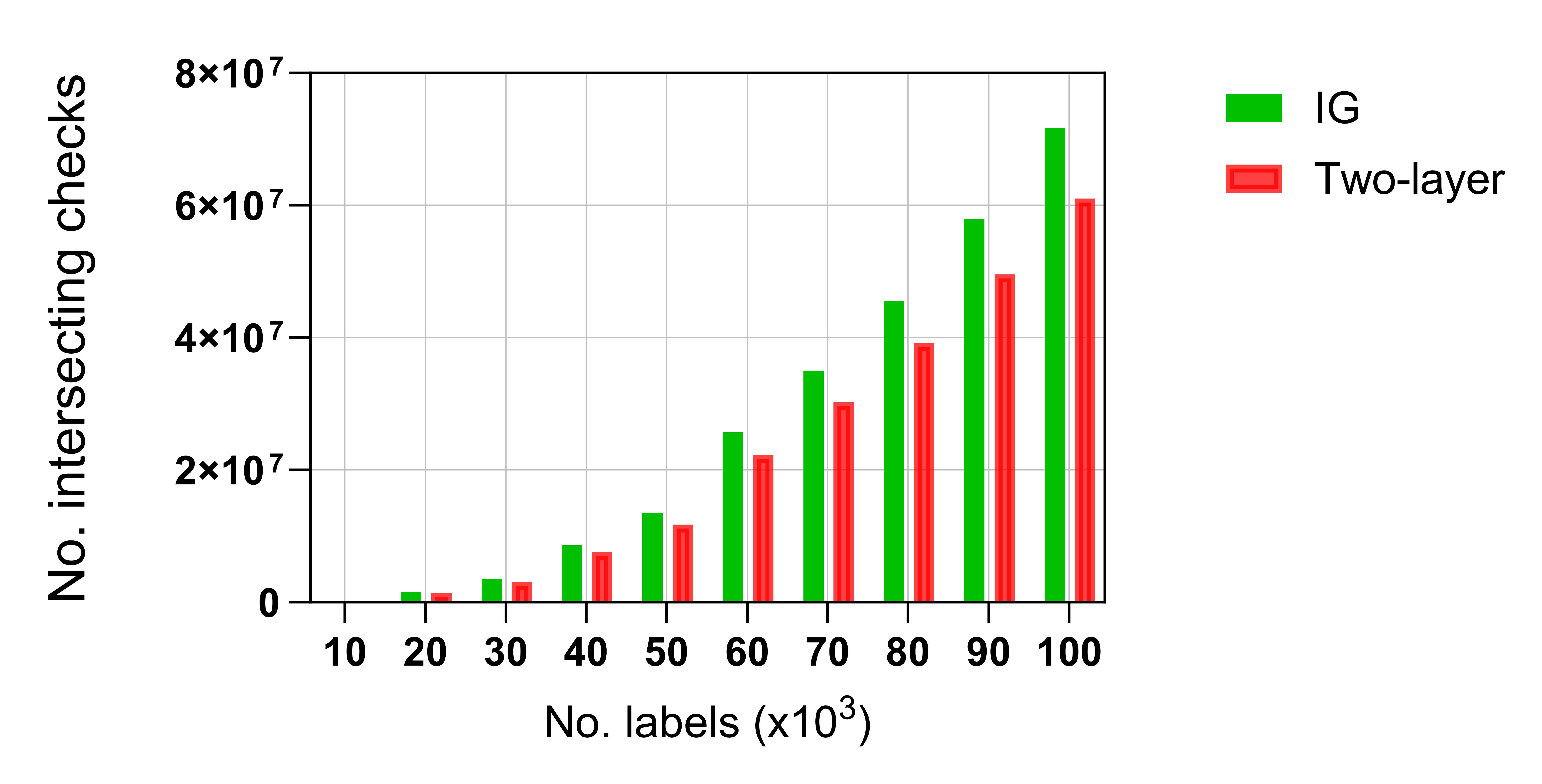}
        \caption{Comparison of the number of intersecting checks}
        \vspace{-0.4cm}
        \label{fig:grh_intersect}%
    \end{figure}

    The superiority of the proposed approach is attributed to its reduction in the computation of intersecting tests by minimizing duplicate results compared to IG. Figure \ref{fig:grh_intersect} compares the number of intersecting tests between the two grid-based methods. As shown in the graph, the gap between these two methods widens as the dataset size increases, resulting in reduced processing time. Additionally, Fig \ref{fig:grh_intersect} illustrates a comparison of the number of intersecting tests between the two grid-based methods. The gap between these two methods widens as the dataset size increases, ultimately leading to reduced processing time. 

\section{Conclusion}\label{sec:conclusion}
    This paper has proposed an efficient implementation for label partitioning in the scalable GLMB filter. By employing a secondary classifying technique, our approach eliminates unnecessary comparisons, resulting in improved computational efficiency. To assess its effectiveness, we conducted simulation experiments, which demonstrated its superior performance compared to other methods, especially as the cardinality of objects increases. Future work includes efficient parallel GLMB filtering on various real-world datasets.

\bibliographystyle{IEEEtran}
\balance
\bibliography{IEEEabrv, ref}

\end{document}